\documentclass[global,twocolumn]{svjour}
\usepackage{graphicx}
\usepackage{dcolumn}
\usepackage{bm}
\usepackage{txfonts}

\begin{document}


\title{Optimized loading of an optical dipole trap for the production of Chromium BECs.}
\author{G. Bismut $^{1,2}$, B. Pasquiou $^{1,2}$, D. Ciampini $^3$, B. Laburthe-Tolra $^{1,2}$,
 E. Mar\'echal $^{1,2}$, L. Vernac $^{1,2}$, and O. Gorceix
$^{1,2}$} \journalname{}
\date{\today}

\institute{ $^1$ Universit\'e Paris 13, Laboratoire de Physique des
Lasers, Institut Galil\'ee, 99 Ave J.B. Cl\'ement, F-93430
Villetaneuse, France,$^2$ CNRS, UMR 7538, 99 Ave J.B. Cl\'ement,
F-93430 Villetaneuse, France, $^3$ CNISM UdR Pisa, Dipartimento di
Fisica E. Fermi, Universita di Pisa , Largo Pontecorvo 3, I-56127
Pisa, Italy}

\maketitle

 \email{etienne.marechal@univ-paris13.fr}

\begin{abstract}

We report on a strategy to maximize the number of chromium atoms
transferred from a magneto-optical trap into an optical trap through
accumulation in metastable states via strong optical pumping. We
analyse how the number of atoms in a chromium Bose Einstein
condensate can be raised by a proper handling of the metastable
state populations. Four laser diodes have been implemented to
address the four levels that are populated during the MOT phase. The
individual importance of each state is specified. To stabilize two
of our laser diode, we have developed a simple ultrastable passive
reference cavity whose long term stability is better than 1 MHz.


\end{abstract}



\section{\label{Intro}Introduction}

Ultracold atoms have found a large amount of applications ranging
from Bose-Einstein condensation (BEC) to exquisitely accurate
measurements. The achievement of BEC in ultracold dilute gases
triggered a burst of activity devoted to the experimental
investigation of various fundamental issues in quantum physics.
Numerous experimental setups are operated to enlarge the field of
applications of trapped ultracold atoms \cite{siteatomtrap}. Most of
these experiments are performed with samples of alkali atoms that
are relatively easy to cool down to the micro-Kelvin range in
magneto-optical traps (MOTs), and that can be evaporatively cooled
down to the nanoKelvin range after their transfer into a magnetic
trap (MT). The extension of the cooling techniques to other atoms is
attractive for several reasons among which one can quote only an
arbitrary selection while, of course, some atoms are pertaining to
several classes of applications or scientific goals. Several atoms
possessing more complex internal level structure than alkalis are
appealing for optical frequency precision metrology (earth-alkali
atoms like Calcium \cite{kraftBECAlkalineEarth:130401}, Strontium
\cite{StellmerBECStrontium:0910.0634v1} and Mercury
\cite{petersenMercury:183004}) or for testing ultracold collision
theories \cite{wienerCollisionsRevModPhys.71.1}. Earth alkali atoms
have also been proposed as quantum registers \cite{Zoller}. Atoms
that come along with a diversity of naturally abundant isotopes
(such as Ytterbium and Chromium) are appealing in the prospect of
producing and studying mixtures of fermionic and/or bosonic quantum
degenerate gases \cite{fukuharaYtterbium:021601}. Furthermore, group
III atoms (like Aluminium, Gallium and Indium
\cite{RehsegroupIII:ApplPhysB70_657}) as well as Iron
\cite{SmeetsIron:ApplPhysB80_833}, have been optically manipulated
in view of potential applications in nanofabrication. Finally,
cooling of highly magnetic atoms opens the possibility to study the
effect of dipole-dipole interactions in degenerated quantum gases.
Among these atoms, Chromium atoms carry a magnetic moment of 6
$\mu_B$, Erbium atoms of 7 $\mu_B$ \cite{berglundErbium:113002} and
Dysprosium, recently laser cooled in a MOT \cite{Dysprosium}, of 10
$\mu_B$. Chromium BECs have been produced during the last five years
in the group of T.Pfau at the Stuttgart University and in our group
\cite{PfauBec,BECParis13}. This article is devoted to the
optimization of the production of such a BEC using an all-optical
procedure. Even though we concentrate on our specific atomic
species, the techniques that we present to reach degeneracy and
raise the number of condensed atoms, apply to many other
experimental endeavors in which the handling of leaky transitions,
the accumulation in metastable states and the need for an optimized
loading into an optical trap are technical bottlenecks that need to
be overcome.

In this paper, we precisely describe the loading of an optical
dipole trap directly from a MOT. Metastable states are populated in
a MOT by optical pumping, and are trapped in a single beam
horizontal optical dipole trap. Thus, cooling in the MOT and
trapping in metastable states become independent. This reduces
light-assisted density-limiting processes, and allows for a larger
density in the optical dipole trap, in the spirit of the dark spot
physics \cite{darkspot}. We describe the optimization of the loading
process by use of  optical depumping towards four different
metastable states. We find that the total number of atoms
accumulated in metastable states is relatively independent of the
metastable states in which the depumping is achieved, but most
sensitive to the depumping rate. We find that optimum loading arises
when the depumping timescale is comparable to the equilibration (or
damping) time in the MOT. Such a strong depumping depletes the MOT
atom number so that light assisted losses (particularly strong in
the case of Cr) becomes negligible and do not limit the loading
efficiency anymore.

Another beneficial consequence of the strong depumping, is that by
limiting the MOT density, it limits the rescattering of light,
leading to MOTs of small size which nicely match the size of the
optical dipole trap. This leads to a very good transfer rate of
atoms from the MOT to the dipole trap, which is very hard to achieve
with MOTs having a larger size. We also discuss the impact of this
optimization on the BEC production. The Appendix describes an
ultra-stable Fabry-Perot cavity, with very good long term stability,
which we use to lock two laser diodes repumping two of the four
metastable states involved.

\section{\label{history}Strategies for the creation of a Chromium BEC}

Reaching quantum degenerate regime with chromium atoms is difficult,
due to unfavorable collisional properties. Specific strategies had
to be invented to tackle with particular magnetic and spectroscopic
properties of chromium. An important particularity of chromium is
that inelastic light assisted collision rates are abnormally high.
As a consequence, chromium magneto-optical trap clouds are small,
both in size and atom number, limited to less than $10^7$ atoms. For
example, the first MOTs to be obtained
\cite{MOT1,MOTChrome2,MOTCr53} were limited to $10^6$ atoms. As
shown in Fig. \ref{FigNiveauEnergie}, laser cooling is realized
using the almost closed strong $^7S_3\rightarrow {^7P_4}$ transition
at 425 nm. Atoms in the excited $^7P_4$ state can decay to the
$^5D_4$ and $^5D_3$ metastable states with a branching ratio of the
order of $10^{-5}$. A natural idea is to plug these leaks by shining
two lasers resonant with the transitions between these metastable
states and the $^7P_4$ or $^7P_3$ state (due to stronger transition
strengths, we repump the metastable atoms via the $^7P_3$ state).
Even with these two repumpers on, the atom number in the MOT remains
limited to a few $10^6$ atoms because of light assisted collisions.
Strategies to produce chromium BEC have to deal with this relatively
small number of atoms in the MOT.

The metastable $^5D$ levels offer a means to prepare large samples
of cold atoms from a MOT. Indeed, as shown in figure 1 and stated
above, atoms in $^7P_4$ state can decay to metastable levels $^5D_4$
and $^5D_3$. If their internal magnetic number $m_j$ is positive
(i.e. if they are in a low field seeking state), they remain trapped
by the magnetic field gradient produced by the MOT coils, while they
are immune to the MOT light. In addition, as the magnetic moment of
chromium atoms is as high as 6 $\muup_B$ for the fully stretched
states, chromium atoms are supported against gravity by moderate
magnetic gradients compatible with the optimal MOT working point.

This continuous loading of the magnetic trap with metastable atoms
is the first key step in the road leading to the achievement of the
first chromium BEC \cite{PfauBec,PfauBecAPB}. From this point, the
strategy of the Stuttgart group is the following: they load a few
$10^8$ chromium atoms in a Ioffe-Pritchard (IP) magnetic trap, using
only the $^5D_4$ level as a reservoir, the $^5D_3$ state being
continuously repumped with a laser resonant with the $^5D_3
\rightarrow {^7P_3}$ transition. This IP magnetic trap is loaded
from a 2D-MOT trap that shares the same magnetic potential. After a
Doppler cooling stage, necessary to increase the initial phase-space
density, and a first evaporation stage in the magnetic trap, atoms
are transferred into a dipole trap. Once the dipole trap is loaded,
atoms are optically pumped into the lowest energy level $(^7S_3,
m_j=-3)$, to overcome the inelastic dipolar relaxation processes
\cite{relaxationdipolaire}, and the final evaporation ramp is
performed in a $1D$ dipole trap and then in a crossed dipole trap.
The whole evaporation procedure cannot be performed entirely in a
magnetic trap because chromium exhibits large dipole-dipole
inelastic collision rates that preclude the BEC formation
\cite{relaxationdipolaire}. The dipole-dipole collisions lead to
relaxation of the initially magnetically trapped low field seeking
state $(m_j=+3)$ toward state of lower energy, which releases
kinetic energy. To avoid this mechanism, atoms have to be trapped in
the true fundamental state $(m_j=-3)$, which cannot be trapped in a
magnetic trap, so that a dipole trap has to be used.

In Villetaneuse, we follow a shortcut by directly loading a dipole
trap with metastable atoms from the MOT, to bypass the intermediate
loading stage of atoms into a Ioffe-Pritchard type magnetic trap. We
have demonstrated that a one beam dipole trap can be efficiently
loaded during the MOT phase \cite{AccumulationEPJD}. For this setup,
an horizontal dipole trap is focused at the position of a
conventional MOT (comprising a quadrupolar magnetic field) so that
metastable atoms accumulate in the combined magnetic plus dipole
trap: atoms are trapped by dipolar forces transversally and by
magnetic forces along the direction of the dipole trap beam. Our
strategy to reach the quantum degeneracy involves, as a crucial
point, the maximization of the number of atoms accumulated in this
combined trap.

A first original idea was to cancel out the effect of the magnetic
trap on metastable atoms during the accumulation stage using
rf-sweeps. We have demonstrated that in presence of appropriate
rf-sweeps, metastable atoms produced during the MOT phase accumulate
in a \emph{pure} dipole trap. The MOT is unaffected by the presence
of the strong rf fields used during the rf-sweeps. This scheme
allows to trap both the high-field and low-field seeker atoms in the
dipole trap, and decreases the confinement along the trapping laser
beam propagation axis. We have shown that this rf sweep scheme
increases the number of atoms loaded in the dipole trap by 50 \%
\cite{sweepRF}. A second decisive step was to force new leaks
towards metastable levels by using a depumping laser, to increase
the number of atoms initially loaded in the dipole trap. In
particular, the accumulation in a new metastable state, $^5S_2$,
turned out to be a decisive step to reach BEC \cite{BECParis13}. In
the present paper, we quantify in details the contributions of the
different metastable levels. This allows us to optimize the atom
number in the BEC, using a procedure that enables the independent
addressing of each of the metastable states that can be loaded from
the MOT into the optical trap.

\section{\label{metastable levels}Metastable level scheme}

\begin{figure}[h]
\centering
\includegraphics[width=3in]{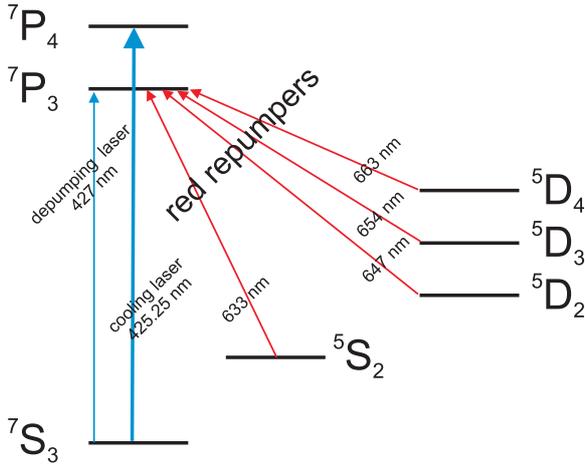}
\caption{\setlength{\baselineskip}{6pt} {\protect\scriptsize
Simplified Chromium energy scheme, showing the relevant levels and
transitions. The strong $^7S_3 \rightarrow {^7P_4}$ transition at
425 nm is used for cooling. The $^7S_3 \rightarrow {^7P_3}$
transition at 427 nm is used for depumping, with a power of 80
$\muup$W. Atoms accumulated in the four metastable states $^5D_4,
{^5D_3}, {^5D_2}$ and $^5S_2$ can be repumped into the fundamental
state using four independent red laser beams (with wavelengths
between 633 and 663 nm.)}} \label{FigNiveauEnergie}
\end{figure}

As shown in Fig. \ref{FigNiveauEnergie}, metastable states have to
be taken into account when cooling Chromium. As already mentioned,
when atoms are cooled using the $^7S_3 \rightarrow {^7P_4}$
transition at 425.6 nm, metastable states $^5D_4$ and $^5D_3$ are
populated due to a weak coupling between these two states and the
excited state $^7P_4$. In addition, by exciting the $^7P_3$ state
with a 'depumping' laser at 427.6 nm, two further metastable states
become also populated : $^5D_2$ and $^5S_2$.  The couplings between
these four metastable states and the two excited states $^7P_4$ and
$^7P_3$ are listed in Tab. \ref{couplingtab}. These couplings remain
small compared to the ones between the $^7S_3$ state and the two
excited states $^7P_4$ and $^7P_3$ which are characterized by a
transition rate $A=3.1 \, 10^7 s^{-1}$ \cite{Nist}. When only the
main transition at 425 nm is excited, the branching ratio between
transitions to the $^5D_4$ state and the $^7S_3$ state is of $4.1 \,
10^{-6}$. For this reason atoms have time to be efficiently cooled
in the MOT before they decay into the metastable states. Coupling to
the $^5S_2$ state is stronger than to the other metastable states,
and atoms can be accumulated mostly in this state, in presence of
the depumping laser, as demonstrated by the reported below
experimental studies.

\begin{table}[htbp]\centering
\begin{tabular}{|l|c|c|c|c|c|}
  \hline
 & $^7S_3$ & $^5S_2$& $^5D_4$ & $^5D_3$ & $^5D_2$ \\
  \hline
 $^7P_4$ & $3.15 \, 10^7$ & forbidden & $127$ & $42$ & \footnotesize{forbidden }\\
\hline
$^7P_3$ & $3.07 \, 10^7$ & $2.9 \, 10^4$& $6 \, 10^3$ & \footnotesize{unknown } & \footnotesize{unknown } \\
\hline
\end{tabular}
\caption{\setlength{\baselineskip}{6pt} {\protect\scriptsize
Transition probabilities in $s^{-1}$ (from the NIST Atomic and
Spectra Database \cite{Nist}, and from
\cite{PfauContinuousLoadingQuad}).}} \label{couplingtab}
\end{table}

\section{\label{loadingtrap}Loading of a dipole trap with metastable atoms}

\subsection{\label{experimental setup}Experimental setup}

Our setup to implement a dipole trap has been already presented in
\cite{BECParis13,AccumulationEPJD}. An horizontal retroreflected
infra-red (IR) laser beam (35 W max, $\lambda=1075$ nm) is focused
at the MOT position with a waist of 42 $\muup$m. This far red
detuned laser realizes a dipole trap into which cold metastable
atoms accumulate. The four metastable levels can be addressed using
four independent red laser beams (see Fig. \ref{FigNiveauEnergie}),
allowing to repump each metastable state through the $^7P_3$ excited
state. We give in the appendix the technical details in relation to
the frequency stabilization of these four repumping lasers. At the
MOT position, these red laser beams have a power of typically 5 mW
and a 1 mm $1/e^2$ radius. The sequence of accumulation is the
following: at t=0, the MOT laser at 425 nm, the depumping laser at
427 nm and the trapping IR laser are switched on. For all these
experiments, the IR laser power was set to 30 W. In addition, during
the accumulation sequence, the rf-sweeps are always switched on. As
explained above, metastable atoms accumulate therefore in a one-beam
pure dipole trap \cite{sweepRF}. As we have developed a laser setup
that can address independently each metastable state, we can choose
to accumulate into one single metastable state or into different
states simultaneously by switching on or off the red laser
addressing the corresponding state during the accumulation stage, as
explained in the next subsection. When a given red laser is on, the
leak to the corresponding metastable state is 'plugged'. After a
given accumulation time, the MOT magnetic field, the MOT beams and
the depumping 427 nm laser are switched off. The four red laser
beams are switched on during 30 ms to repump all the metastable
atoms into the fundamental state $^7S_3$. Atoms are then polarized
in the $m_j=-3$ state with a $\sigma^{-}$ pulse resonant with the
$^7S_3 \rightarrow {^7P_3}$ transition at 427 nm, in presence of a
bias field of about 2 Gauss. Finally, the optical trap is switched
off, and after a time of flight of typically 1 ms, the atom cloud is
imaged by absorption imaging using a laser circularly polarized and
resonant with the cooling transition at 425 nm.

\subsection{\label{different reservoirs}Simultaneous or independent loading of the different reservoirs}
\begin{figure}[h]
\centering
\includegraphics[width=3in]{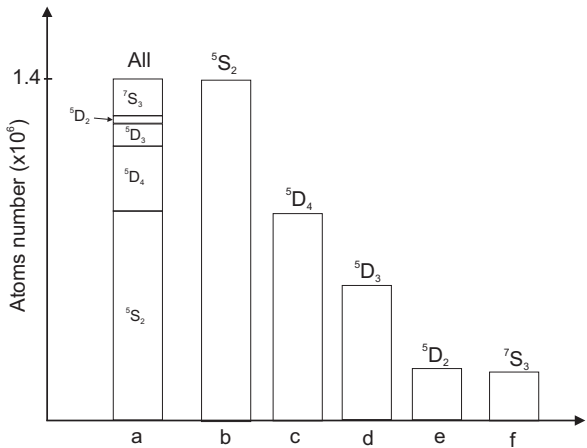}
\caption{\setlength{\baselineskip}{6pt} {\protect\scriptsize Numbers
of atoms in the dipole trap, after the accumulation and the
repumping in the fundamental state $^7S_3$. (a) simultaneous
accumulation in all metastable states, and state composition.
(b)-(e) : accumulation in a single reservoir, the other reservoirs
being plugged during the accumulation phase. (f) Accumulation only
in $^7S_3$ when all the reservoirs are plugged.
}} \label{FigureHistogrammes}
\end{figure}

We have studied the influence of the different repumpers on the
accumulation of metastable atoms in the trap. A naive idea is that
the number of accumulated atoms should raise if we increase the
number of reservoirs. Nevertheless, for a given reservoir, loaded
with only one given state, the maximum number of atoms in this
reservoir is fixed by an equilibrium between the loading rate and
inelastic collision rates. The loading rate is related to the
transition probabilities tabulated in Tab. \ref{couplingtab} and to
the power of the cooling and depumping lasers, which sets the
excited state population in $^7P_3$ and in $^7P_4$. Inelastic
collisions which limit the number of trapped metastable atoms are of
two types: light assisted inelastic collisions with $^7P_4$ and
$^7P_3$ atoms and 2-body inelastic collisions between metastable
atoms. When several reservoirs are loaded simultaneously, 2-body
inelastic collisions between the different metastable states also
occur. The optimization problem at stake is therefore complex, the
number of parameters being large, and the interspecies inelastic
collision rates between the metastable states being unknown. In Fig.
\ref{FigureHistogrammes} we show the comparison between the final
number of atoms that can be loaded in the dipole trap when all the
metastable states are populated during the accumulation phase
(column a), and when only one of these reservoirs is filled during
the accumulation phase, the other metastable states being
continuously repumped (columns b to e). We obtain a surprising
result : as shown by the first two columns, when one accumulates
atoms in all the metastable states (column $\mathbf{a}$) or only in
the $^5S_2$ state (column $\mathbf{b}$), the same final atom number
is reached within the experimental error bars ($\approx$ 10 \%). In
other words, we obtain the same final result if we repump the
$^5D_4$, $^5D_3$ and $^5D_2$ levels continuously or if we repump
them only at the end of the accumulation phase thus effectively
accumulating atoms in these states. The $^5S_2$ state is therefore a
reservoir that allows simply by itself to optimize the trap loading,
and is more favorable than any of the metastable $^5D$ states. We
will therefore focus on this particular state in the next two
paragraphs. We also show on column $\mathbf{a}$ the population
distribution of the different metastable states at the end of the
accumulation, when the four channels are open. To perform this
measurement, we have repumped only the measured channel at the end
of the accumulation phase; the number of atoms accumulated in the
ground state  $^7S_3$ is obtained when we do not repump any
metastable states. We observe that atoms are mainly loaded in
$^5S_2$, $^5D_4$ state and also in the fundamental state $^7S_3$.
Comparing columns $\mathbf{b}$ to $\mathbf{e}$  with column
$\mathbf{a}$ show that for a given state, the number of accumulated
atoms is maximized when the other states are not populated. If not,
inelastic collisions with other metastable states increase the
losses and lead to a smaller steady state atom number. Finally, we
obtain that only very few atoms are accumulated in the $^5D_2$ state
(column $\mathbf{e}$). Indeed when all repumpers are on (column
$\mathbf{f}$), we obtain almost as many atoms as when all but the
647 nm repumpers are on. We therefore estimate that only $3.10^4$
atoms are accumulated in the $^5D_2$ state to be compared to the
$2.10^5$ atoms  accumulated directly in $^7S_3$. This indicates that
the coupling rate between $^5D_2$ and $^7P_3$ is very small.

A simple theoretical model can account for these experimental
findings. We will show here that the experimental outcomes can be
explained by assuming that depumping in metastable states is
achieved in a timescale which is short compared to the timescale for
light assisted losses in the MOT (as will be checked in next
paragraph). We will also assume that the inelastic loss parameters
for all different combinations of metastable states are all roughly
equal. With these  two only assumptions, simple rate equations
account for the observation that accumulating only in  the $^5S_2$
states while continuously repumping all other metastable states
(column $\mathbf{b}$) leads to the same number of accumulated atoms
than accumulating in all metastable states (column $\mathbf{a}$).


For example, in case there are two metastable states, the rate
equations read
\begin{eqnarray}
\frac{dN_1}{dt}=N_{MOT} \Gamma_1 - \beta \frac{N_1}{V}N_1 - \beta \frac{N_2}{V} N_1 \\
\frac{dN_2}{dt}=N_{MOT} \Gamma_2 - \beta \frac{N_2}{V}N_2 - \beta \frac{N_1}{V} N_2 \\
\frac{dN_{MOT}}{dt}=\Gamma - N_{Mot} (\Gamma_1+\Gamma_2)
\end{eqnarray}
where $N_{1,2,MOT}$ is the number of metastable atoms in states 1, 2
or in the MOT, $\beta$ the inelastic loss parameter in metastable
states, and $V$ the effective trap volume. $\Gamma_{1,2}$ are the
depumping rates into states 1 and 2. $\Gamma$ is the loading rate of
the MOT. Note that, when the light assisted collisions are not
negligible, a new term has to be added to equations 1 to 3 of the
form $\beta_{P-i}n^*_{MOT}N_i$  where $n^*_{MOT}$ is the density of
excited $^7P_4$ atoms, and $\beta_{P-i}$ is the the light assisted
loss rate between atoms in  $^7P_4$  state and atoms in state 1, 2
or the MOT atoms.

We solve this set of  simple equations in two different cases. In
the first case, we assume that metastable atoms $2$ are continuously
repumped, so that atoms only accumulate in state 1. The steady state
for the MOT atom number is $N_{MOT}=\frac{\Gamma}{\Gamma_1}$, and
the steady state for the metastable atom number is
$N_1=\sqrt{\frac{V \Gamma}{\beta}}$. In the second case, we
continuously accumulate atoms in both metastable sates,
corresponding to a steady state situation
$N_{MOT}=\frac{\Gamma}{\Gamma_1 + \Gamma_2}$ and
$N_1+N_2=\sqrt{\frac{V \Gamma}{\beta}}$.

This simple model (which can be easily extended to include the four
metastable states) hence reproduces our main experimental
observations: accumulating in all metastable states or in only the
$^5S_2$ state (by continuously repumping all other metastable
states) leads to the same final atom number. The reason for that is
that depumping  to the $^5S_2$ is so fast that inelastic losses can
be neglected in the ground state, so that repumping all metastable
states but one results in optical pumping to the state which is not
repumped. The qualitative agreement between our observations and our
simple model also indicates that the inelastic loss parameters
corresponding to the different combinations of metastable atoms are
of the same order of magnitude. The experimental results show
nevertheless that, when we do not accumulate atoms in the $^5S_2$
state by repumping continuously this level with a repumper at 633
nm, we accumulate less atoms (columns c,d e) in other states. This
shows the limitations of our simple model: first, the 2-body loss
rate parameter for $^5D_4$ state is twice bigger than for $^5S_2$
(as measured in subsection \ref{inelastic collisions}) leading to
finally less atoms accumulated in this state (column c). Second, the
depumping rates to the $^5D_3$ and $^5D_2$ states are very small so
that in addition light-assisted  collisions are no more fully
negligible when we accumulate only in these states (column c and e).

\subsection{\label{5S2 loading}Optimization of the $^5S_2$ loading}

\begin{figure}[h]
\centering
\includegraphics[width=3in]{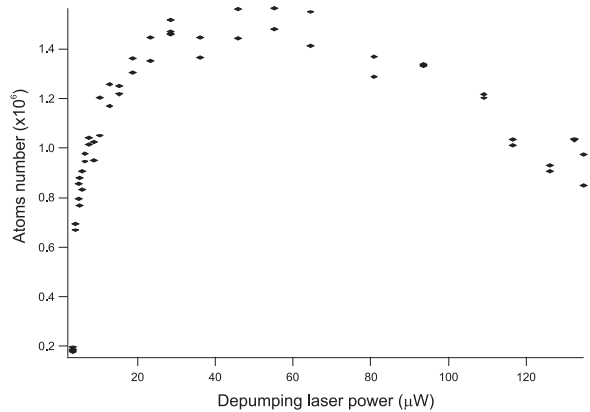}
\caption{\setlength{\baselineskip}{6pt} {\protect\scriptsize
Optimization of the number of atoms accumulated in the $^5S_2$ state
in the dipole trap as a function of the depumping laser power. The
atom number is measured after repumping in the fundamental  $^7S_3$
state. The initial non zero value when the repumper is off
corresponds to atoms directly loaded in the dipole trap from the MOT
in the $^7S_3$ state.}} \label{FigOptimisation5S2}
\end{figure}

In Fig. \ref{FigOptimisation5S2}, we present the number of atoms
accumulated in the dipole trap when only the $^5S_2$ state is loaded
during the accumulation phase as a function of the power of the
depumping laser (at 427 nm). This laser is circularly polarized, and
is collimated with a $1/e^2$ radius of 1.5 mm at the output of an
optical fiber and shone to the MOT. We observe a plateau of optimal
values between 30 and 70 $\muup$W. If the depumping power is too
small, there are few atoms in the excited $^7P_3$ state and the
loading rate to the $^5S_2$ reservoir is too weak. The number of
atoms increases with the depumping laser power until the saturation
is reached. If the power is too strong, another phenomenon occurs.
The MOT is affected by the depumping laser: we observe a decrease of
the MOT fluorescence when the depumping laser power is raised. The
optimum loading is reached when atoms have time to be cooled by the
MOT before decaying into the metastable states. To estimate the
cooling time in the MOT, $T_{MOT}$, we first accumulate atoms in
metastable states in a magnetic trap (the dipole trap is
switched-off). We then switch-on the MOT laser beams with the
repumping lasers for an adjustable time and we measure the time
evolution of the initially large volume of the magnetic trap toward
the steady state MOT volume. We obtain a $1/e$ time evolution
$T_{MOT}$ = 10 ms. This time has to be compared to the time
$T_{depump}$ spent by atoms cycling in the MOT before being
depumped. To measure $T_{depump}$, we monitor the MOT fluorescence
change when we switch on the depumping laser. The MOT fluorescence
steady level evolved within a time scale $T_{depump}$ to a slightly
lower value. For a power of 35 $\muup$W we measure a 1/e time
constant of $T_{depump}$ = 5 ms. These results confirm that the
optimal loading is obtained when $T_{depump}$ and $T_{MOT}$ are of
the same order of magnitude.

We also compare the depumping time $T_{depump}$ to the time
associated to light assisted collisions $T_{Light}$. We estimate
this time by measuring the MOT lifetime, when all the repumpers are
ON, and when we suddenly switch off the Zeeman slower beam. We
measure $T_{Light}$=14 ms. In presence of the depumping beam, the
MOT fluorescence is decreased by a factor of two so that we can
estimate that  $T_{Light}\approx$30 ms. This time is six time larger
than $T_{depump}$ confirming that light assisted collisions are not
limiting the loading of the dipole trap when accumulating in
$^5S_2$. Note however that if we accumulate in $^5D_{4,3,2}$, the
timescale for depumping is raised by a factor of at least five so
that light assisted collisions may then play a role and limit the
atom number.

\subsection{\label{inelastic collisions}Inelastic collisions in the $^5S_2$ state}
The $^5S_2$ state is a particularly interesting reservoir, because
our findings show that the loading is almost optimal when atoms
accumulate only in this state. The loading rate through the coupling
with the $^7P_3$ state is five times bigger for this state than for
the $^5D_4$ state (Tab. \ref{couplingtab}, \cite{Nist}) which favors
the accumulation in this state. In the dipole trap, the number of
accumulated atoms is limited by inelastic collision rates. We have
therefore measured the 2-body inelastic collision rate $\beta$ of
this state. For this, we measured the decay of the $^5S_2$ atom
number: once we have switched off the depumping and the MOT laser
beams, we repump the $^5S_2$ state and take an absorption picture
after a given time. During the decay, the density variation is
locally governed by the following equation:

\begin{equation}\label{equ1}
\frac{dn(\textbf{r},t)}{dt}=-\Gamma n(\textbf{r},t) - \beta
n(\textbf{r},t)^2
\end{equation}

where $\Gamma$ characterizes the one-body losses.  In the present
case, one body losses are mostly due to collisions with hot atoms
coming from the thermal chromium beam, while in presence of the MOT
beam light assisted collisions between atoms contribute also to the
1-body loss rate. We have checked experimentally that the shape of
the atom density is well fitted by a Gaussian distribution,
independent of time
$n(r,t)=\frac{N(t)}{V_0}e^{-(\frac{x^2}{x_0^2}+\frac{y^2}{y_0^2}+\frac{z^2}{z_0^2})}$
where $V_0=x_0y_0z_0\pi^{3/2}$. In that case the solution of Eq.
\ref{equ1} can be integrated to give the evolution equation of the
total atom number $N(t)$:

\begin{equation}
\frac{dN(t)}{dt}=-\Gamma N(t)-\frac{\beta}{2^{3/2}V_0}N(t)^2
\end{equation}

\begin{figure}[h] \centering
\includegraphics[width=3in]{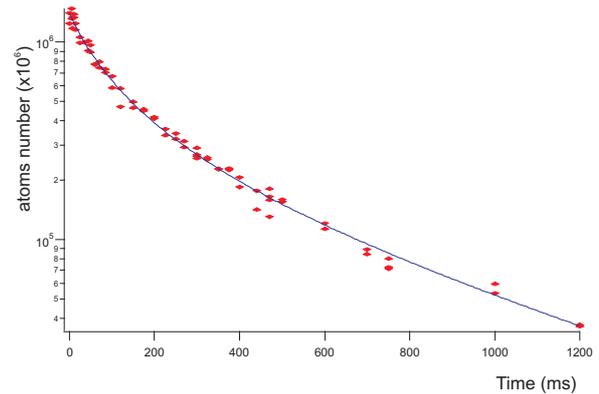}
\caption{\setlength{\baselineskip}{6pt} {\protect\scriptsize Decay
curve for the number of trapped $^5S_2$ atoms after the MOT and
repumper beams have been switched off. The fit using Eq. \ref{equ3}
yields $\beta=1,32.10^{-11}$ cm$^3$.s$^{-1}$ for the two-body loss
coefficient (full line).}} \label{Figdecay5S2}
\end{figure}

\noindent The solution of this equation is given by :

\begin{equation}\label{equ3}
N(t)=\frac{N_0e^{-\Gamma t}}{1+\frac{\alpha
N_0}{\Gamma}(1-e^{-\Gamma t})}
\end{equation}

where $\alpha=\frac{\beta}{2^{3/2}V_0}$ and $N_0=N(t=0)$. The
observed non exponential decay is presented in Fig.
\ref{Figdecay5S2}. The fit to the data using Eq. \ref{equ3} yields
$\alpha$ , $N_0$ and $\Gamma$. The value of $\beta$ is related to
$\alpha$ with the relation $\beta=2^{3/2}\alpha V_0=2^{3/2}\alpha
N_0/n_0$ where $n_0$ is the initial peak density. The direct
measurement of the density by in situ absorption imaging is not
possible as all the light is absorbed at the center of the cloud,
leading to truncated absorption profiles. For this reason, we deduce
the peak atomic density from measurements of the dipole trap beam
parameters and by using the theoritical light shifts. The power of
the optical trap laser is 2 x 30 W at the MOT position, the waist is
$w_0$ = 42 $\muup$m and the estimated light shift is 12.9 MHz for
the $^5S_2$ state. The theoretical calculation of this light shift
is presented in \cite{AccumulationEPJD}. The peak atomic density is
then obtained by substituting a parabolic approximation to the
dipole potential, and by considering a Maxwell Boltzmann
distribution for the atom density in this potential. The temperature
is measured by time of flight. For the conditions of Fig.
\ref{Figdecay5S2} we found a temperature of 82 $\muup$K and we
obtained a peak density of $1,30.10^{12}$ cm$^{-3}$ giving a value
of $\beta = 1,3.10^{-11}$ cm$^3$.s$^{-1}$. With additional
measurements, we finally estimated a value of $\beta=1,6.10^{-11}\pm
0,4.10^{-11}$ cm$^3$.s$^{-1}$. The main error comes from the
estimate of the peak density. Using the same procedure, we measured
the inelastic decay  rate in the $^5D_4$ state to be $3,5.10^{-11}$
cm$^3$.s$^{-1}$. This value is in reasonable agreement with the
already reported value of $2.6.10^{-11}$ cm$^3$.s$^{-1}$ and
$3,3.10^{-11}$ cm$^3$.s$^{-1}$
\cite{PfauContinuousLoadingQuad,accumulationMT}. We therefore
conclude that the collisional properties of the $^5S_2$ state are
slightly more favorable to accumulation than those of the $^5D_4 $
state as $\beta$ is about twice smaller for $^5S_2$.

We also obtain that in the present case, the number of atoms loaded
in the dipole trap is limited by two-body inelastic collisions
between metastable atoms (as in \cite{AccumulationEPJD}). This is
very different from the case of the accumulation of metastable
chromium atoms in a magnetic trap
\cite{PfauContinuousLoadingQuad,PfauContinuousLoadingIP}, where
light assisted collisions between excited atoms $^7P_4$ from the MOT
and
metastable atoms are the dominant limiting mechanisms. 
This fact is clearly confirmed by the following experimental
observation: when we measure the decay of the $^5S_2$ state
presented in Fig. \ref{Figdecay5S2} in presence of the MOT, the
initial decay, i.e. when the density is high, is unchanged. This
also confirms that the simple model presented in subsection 4.2 is
valid.

\subsection{\label{effetsurBEC}Relevance of the different metastable repumpers for the chromium BEC production}

Before the conclusion, we sum up here our accumulation procedure for
the production of the BEC (typically containing  15 to 20 thousand
atoms in a pure condensate). As pointed out in sub-section
\ref{different reservoirs}, two equivalent procedures can be used to
load efficiently the dipole trap, before starting the evaporation
phase: either accumulating atoms in  only $^5S_2$ or accumulating
atoms in the four metastable channels (column a of Fig.
\ref{FigureHistogrammes}). We use this last procedure: the repumping
laser are switched-off during the accumulation phase, and are
switched-on at the end to transfer all metastable atoms in $^7S_3$.
In addition, we add, as explained in \cite{BECParis13}, a 'dark
spot' repumping laser during the accumulation phase, resonant on the
$^5S_2 \rightarrow {^7P_3}$ transition that continuously recycles
$^5S_2$ atoms that are not in the dipole trap volume. This 'dark
spot' laser (which was not used during the studies presented in this
paper) allows to increase the final atom number in the BEC by 10\%.
We can finally estimate the relative importance of the different
repumping lasers. In our case, if we do not repump atoms in the
$^5S_2$ state at the end of the accumulation we obtain no BEC. If we
do not repump $^5D_4$ atoms we decrease the BEC atom number by 78
\%. And if we do not repump $^5D_3$ atoms, we decrease the BEC atom
number by 25 \%. Finally, as already mentioned previously, the
repumping of the $^5D_2$ state has no effect on the final atom
number on the BEC so that we do not use this repumping laser for the
BEC production.

\section{\label{label5}Conclusion}

We have presented here our strategy to optimize the loading of a
dipole trap for the production a chromium BEC. The dipole trap is
directly loaded during the MOT phase and we optimized the loading by
using a controlled depumping toward metastable states. We have shown
that the depumping is strong enough to overcome the light-assisted
limiting processes. The optimal depumping is obtained when the
depumping rate is of the order of the MOT formation (damping) time
$T_{MOT}$. We also focused on the relative relevance of the four
metastable states. We found that the $^5S_2$ state is more favorable
than the other metastable states in view of raising the overall
final number of trapped Cr atoms. We related this to the inelastic
collision rate $\beta$ of this state and compared its value to the
one of the that states. We also found  that the $^5D_2$ state is
only very weakly coupled to $^7P$ states so that it takes no part in
the BEC production.

In the appendix, we have presented a method of stabilizing the
frequency of laser diodes by locking them to a passive ultrastable
Fabry-Perot cavity. The method requires very simple mechanics,
optics and electronics. It could apply to many experiments in the
ultracold atom physics domain.

Acknowledgments: LPL is Unit\'e Mixte (UMR 7538) of CNRS and of
Universit\'e Paris Nord. This research has been supported by
Minist\`ere de l'Enseignement Sup\'erieur et de la Recherche (CPER)
and by IFRAF (Institut Francilien de Recherche sur les Atomes
Froids). G. Bismut acknowledges financial support by IFRAF. We thank
the members of the HOTES group from LPL for giving us the ULE rod
that we used to build the ultrastable cavity.

\newpage

\appendix
\section{\label{paraCavite} Appendix: Description of the passive ultrastable cavity }

The repumping lasers frequencies are generated by four independent
extended cavity laser diodes delivering a power of 10 to 25 mW. To
address the metastable states, each laser frequency has to be
stabilized within 1 MHz to the corresponding chromium atomic
transition.

For this purpose each laser's frequency is locked to an
environmentally isolated Fabry-Perot stable cavity using the Pound
Drever Hall (PDH) technique \cite{PoundDreverHall}. In brief, the
laser beam is first split in two, the main power is used for the
experiment and the other part is used for the lock. The laser
current is gently modulated to a fixed frequency chosen between 10
and 20 MHz, through a bias T connected to the laser diode. This
modulation gives rise to two main sidebands in the laser frequency
spectrum. The reflected light from the Fabry-Perot cavity is
detected on a fast photodiode. The phase sensitive detection at the
modulation frequency gives the PDH dispersive error signal. This
error signal is treated with a standard proportional integral
differential regulator and then used to lock the laser frequency
through the regulation of the laser current and of the piezo voltage
of the diffraction grating of the extended cavity. With the PDH
locking technique, many lasers can be locked on the same cavity,
simply by assigning a different modulation frequency to each laser.
For practical reasons, we have developed two distinct cavities to
lock the four repumping lasers. The main difference between the two
setups lies in the Fabry-Perot reference cavity.

For the first cavity, the cavity's length can be scanned by a PZT
and locked to the $^7S_3 \rightarrow {^7P_4}$ transition of
$^{52}Cr$ by use of saturation spectroscopy. This set-up was
developed in the early stages of the construction of our experience
\cite{MOTCr53} and is used to lock the two diodes at 663 nm ($^5D_4$
level) and at 633 nm ($^5S_2$ level), corresponding to the two most
important transitions. Indeed, with only these two diodes locked, we
can produce a BEC. The Fabry-Perot cavity is first used to frequency
stabilize the 851 nm Ti-Sapphire laser (from Tekhnoscan), pumped by
a Verdi laser (from Coherent). The IR laser light is then frequency
doubled in an external cavity at 425 nm. Finally, the Fabry-Perot
cavity length is locked to the $^7S_3 \rightarrow {^7P_4}$
transition using a saturated absorption in a chromium hollow cathode
lamp. One important drawback of this setup is that the different
feedback loops are highly intricate. If, for example, the
Ti-Sapphire laser or the laser or the doubling cavity  suddenly
unlocks, the Fabry-Perot cavity is then no more referenced and
locking of the repumping lasers has to be carried out again. As we
have added two new red repumpers, we have decided to implement a
passive reference cavity, whose length is therefore fixed in time
and can be used everyday as a robust absolute frequency reference.
We will describe here in more details the passive cavity, which is a
simple versatile system that can be useful to many experiments where
a long term stability is required.


\begin{figure}[h] \centering
\includegraphics[width=3in]{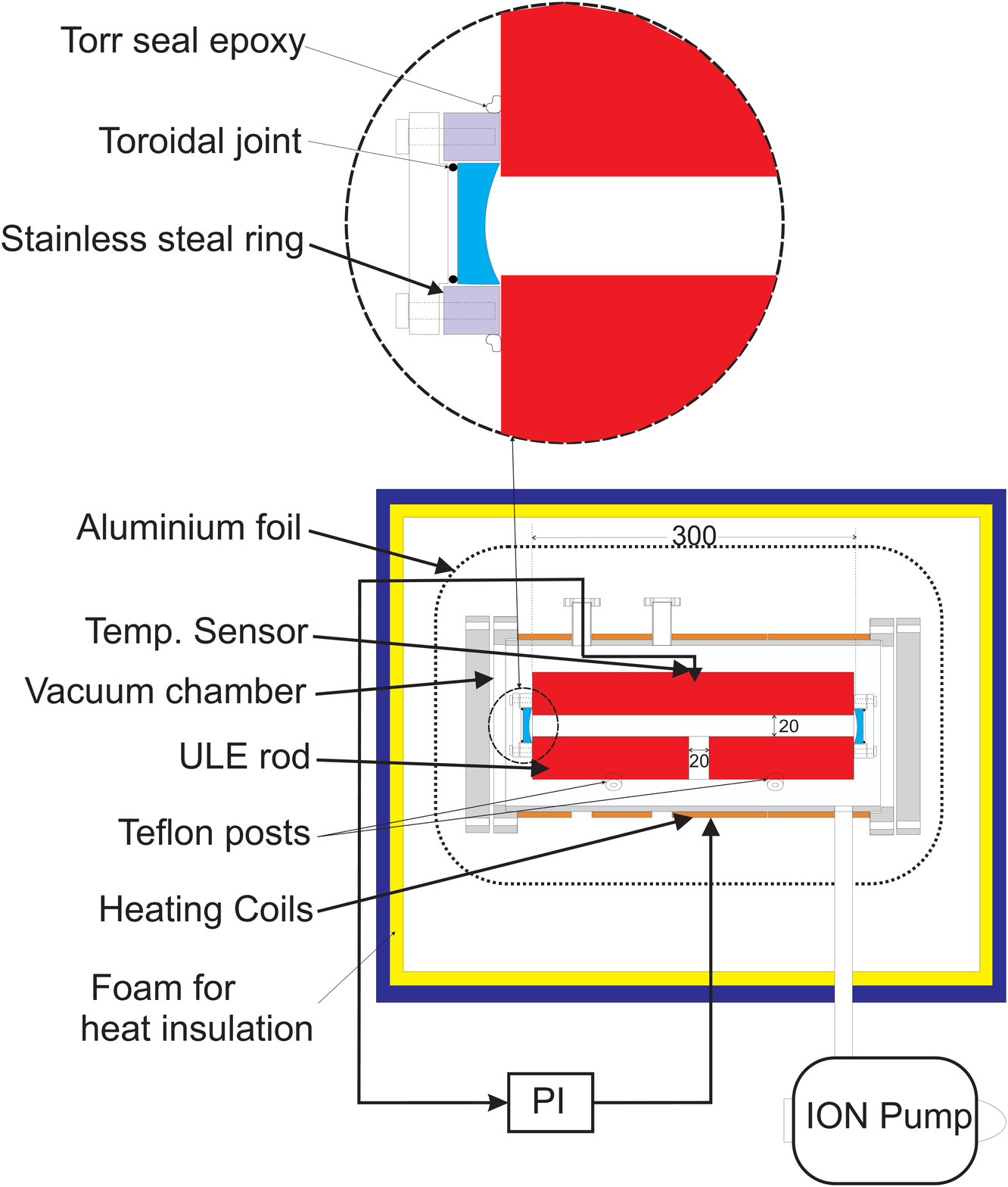}
\caption{\setlength{\baselineskip}{6pt} {\protect\scriptsize
Ultra-stable cavity set-up and detail of the mirrors fixing}}
\label{Cavite}
\end{figure}

The ultra-stable cavity scheme is shown in Fig. \ref{Cavite}. The
cavity is a degenerate confocal cavity, with a measured finesse of
82.  The two converging spherical mirrors (curvature radius of 300
mm) are fixed at the extremities of a Ultra Low Expansion (ULE)
glass spacer, which defines the distance between the two mirrors.
The mirrors fixing is shown in the inset of Fig. \ref{Cavite}. A
first ring (in stainless steel) is glued on the glass using Torr
Seal epoxy . Another ring, screwed on the first, allows to gently
push the mirror, through an helicoflex metallic joint, and put it in
contact with the ULE rod. The ULE spacer (300 mm length, 100 mm
diameter) has an axial aperture of 20 mm diameter for the laser beam
propagation. It is also pierced in its center by a 20 mm hole for
the pumping. The Fabry-Perot cavity is located inside an Ultra-High
vacuum chamber pumped with a 10 l/s  ion pump (P $<$ 10$^{-8}$ mBar
and rests on 4 teflon posts (2 mm diameter). In order to achieve the
required temperature stability, heating coils are located around the
external surface of the vacuum vessel containing the ULE rod. The
heating coils are home made with a copper wire winded up around
cardboard papers (pitch of 5 mm), which are then wrapped up around
the vacuum chamber. This allows to have a uniform repartition of the
heating foils at the vacuum vessel surface. A partially stable
temperature environment is provided to the vacuum vessel by
uniformly covering it with aluminium foils and by housing it in a
thermal insulation box. The addition of the box reduces temperature
fluctuations of the ULE rod by a factor 3. The temperature
fluctuations and drifts are measured by placing several sensors
(thermistors mounted on a Wheatstone bridge) on the surface of the
vacuum vessel and inside it on the ULE rod. One of these sensors,
which is stuck on the surface of the ULE rod, is giving the feedback
signal for our Proportional-Integrator (PI) controller which acts
upon the current sent through the heating coils. The thermal time
constant of the ULE rod relative to the heating coils was measured
to 8 hours, the coupling being mostly due to radiations from the
internal surface of the vacuum vessel. The temperature of the cavity
is set to 26 $^\circ C$, slightly above the room temperature. The
room temperature is regulated to 23 $^\circ C$ with fluctuations
less than $\pm 1^\circ$. The time constant of our PI corrector is
set to 100 s so as to compensate for fast thermal fluctuations. In
all cases it takes around 10 hours for the system to stabilize to a
regime where the temperature at the measurement point on the surface
of the ULE rod varies by less than 500 $\mu K$. This temperature is
measured by monitoring the error signal. We achieve a long term (1
year) stability below 1 MHz , tested using a frequency stabilized
He-Ne laser at 633 nm. Using an AOM, the frequency of the laser
light is swept over a free spectral range of the cavity and the
position of the transmission peak of the laser light in the cavity
is spotted at each frequency sweep. We could also daily check that
the frequency of the cavity is stable, within 1 MHz, compared to the
frequency of the $({^5D_3} \rightarrow {^7P_3})$ repumping chromium
transition at 654 nm.

 During the development phase, we have measured the Coefficient of
 Thermal Expansion (CTE) of the whole cavity by ramping slowly the
 temperature of the ULE rod (3 days from 20 $^\circ C$ to 35 $
 ^\circ C$). Although a temperature dependence of CTE is recorded
 during measurements,  a significant hysteresis on the temperature
 dependence of the cavity length is observed during the ramping up
 and down, which affects greatly the performances of the cavity in
 terms of sensitivity to temperature fluctuations. With our
 observations, a most reasonable assumption is that it stems from
 the frictions at the interface between the cavity and the teflon
 posts. Hysteresis makes every effort to reach a region of lower CTE
 useless. This induces a maximal CTE of $\alpha=4,4.10^{-7} K^{-1}$
 at ambient temperature, which is at least 10 times worse than
 expected from such an ULE \cite{RefHotes}. This relatively high CTE
 could be explained by the thermal expansion of the mirrors fixing
 system. Despite this hysteresis phenomenon, the very good
 performance of the temperature control allows us to reach the long
 term stability that is required. It must be noted that using the
 temperature sensor on the external surface of the vessel as the
 feedback branch for our frequency stabilization system did not
 provide a good enough screening of outside temperature
 fluctuations. Doing so, the frequency fluctuations has been measured
 to be around 6 MHz over 5 hours while the room temperature variation has been
 as large as 1K. We therefore concluded
 that for our setup, it was crucial to put the feedback sensor
 directly on the ULE rod surface.

We point out here that our cavity setup is remarkably simple. It
uses standard electronics and optics (excepted the ULE rod, whose
design is also simple). The cavity finesse is rather low, the
fixation of the mirrors is basic. The obtained performances are far
from the ones of very carefully designed high finesse ultrastable
cavities developed in advanced labs \cite{alnis}. Nevertheless our
scheme for frequency stabilization if sufficient for most of the
cold atoms experiments.

\end{document}